\documentclass[aps,pre,twocolumn,groupedaddress,superscriptaddress, showpacs]{revtex4-1}

\usepackage{graphicx}
\usepackage{textcomp}
\usepackage[usenames,dvipsnames]{color}

\usepackage{transparent}

\usepackage{tipa}

\usepackage{amsmath}

\usepackage{amsfonts}

\usepackage{amssymb}

\usepackage{dcolumn}

\usepackage[english]{babel}

\usepackage{array}

\usepackage{dsfont}

\begin{document}

\title{Cutting self-similar space-filling sphere packings}

\author{D. V. St\"ager}

  \email{staegerd@ethz.ch}

  \affiliation{Computational Physics for Engineering Materials, IfB, ETH Zurich, Wolfgang-Pauli-Strasse 27, CH-8093 Zurich, Switzerland}

\author{H. J. Herrmann}

  \email{hans@ifb.baug.ethz.ch}

  \affiliation{Computational Physics for Engineering Materials, IfB, ETH Zurich, Wolfgang-Pauli-Strasse 27, CH-8093 Zurich, Switzerland}
  
  \affiliation{Departamento de F\'isica, Universidade Federal do Cear\'a, 60451-970 Fortaleza, Cear\'a, Brazil}

\begin{abstract}
Any space-filling packing of spheres can be cut by a plane to obtain a space-filling packing of disks. Here, we deal with space-filling packings generated using inversive geometry leading to exactly self-similar fractal packings. First, we prove that cutting along a random hyperplane leads in general to a packing with a fractal dimension of the one of the uncut packing minus one. Second, we find special cuts which can be constructed themselves by inversive geometry. Such special cuts have specific fractal dimensions, which we demonstrate by cutting a three- and a four-dimensional packing. The increase in the number of found special cuts with respect to a cutoff parameter suggests the existence of infinitely many topologies with distinct fractal dimensions.
\vspace{5mm}

\end{abstract}

\maketitle

\section{Introduction}

\begin{figure}[]
\begin{center}
	\includegraphics[width=\columnwidth]{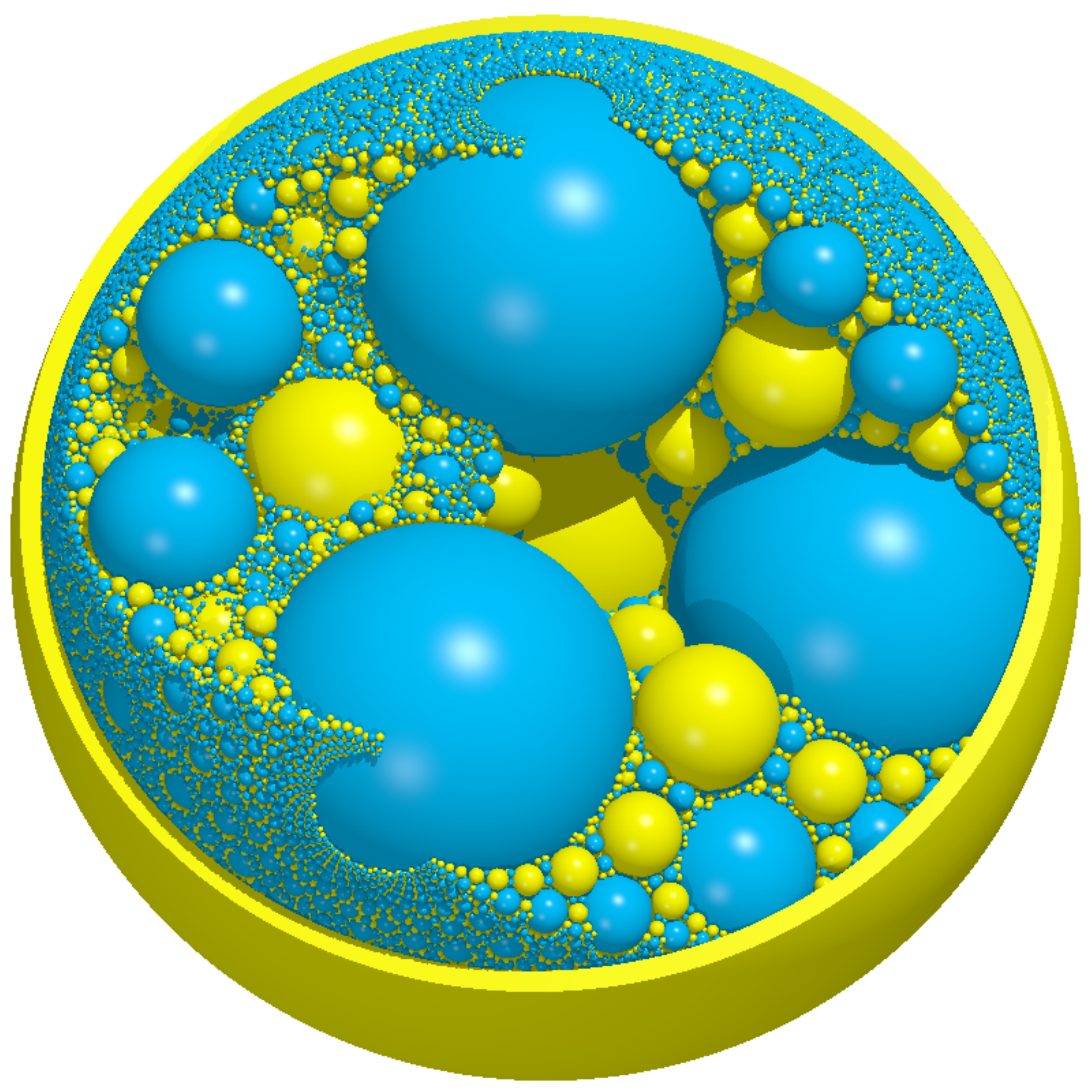}
\end{center}
\caption{
\label{fig:RezaPacking} Self-similar space-filling sphere packing, first discovered by Ref.~\cite{Baram2004a}, constructed using inversive geometry. This particular packing is bipartite, such that one can color the spheres using two colors such that no spheres of same color touch. The packing is enclosed in the unit sphere, which is visualized as a surrounding shell. Spheres with a radius larger than $0.005$ are shown, and some spheres are removed to allow looking inside the packing.
}
\end{figure}

Space-filling sphere packings, as the one shown in Fig.~\ref{fig:RezaPacking}, consist of polydisperse spheres and are completely dense in the limit of infinitesimally small spheres. They can be seen as ideal references for highly dense granular packings, which due to their diverse applications are studied experimentally and theoretically \cite{Ayer1965,Jodrey1985,
Yu1988,Ouchiyama1989,Soppe1990,Konakawa1990,Standish1991,
Yu1993,Anishchik1995,Elliott2002,Sobolev2010,Rahmani2014,Martin2014,Martin2015}. Space-filling packings are fractal, and the size distribution of spheres follows asymptotically a power-law, from which the fractal dimension can be estimated.  

There are many different types of space-filling packings, constructed with different methods, such as the Apollonian Gasket \cite{Kausch1970,Boyd1973,Manna1991,Manna1991a,Borkovec1994,Anishchik1995,Mantica1995,
Doye2005,Varrato2011,Kranz2015} and its generalizations \cite{Bessis1990,Pickover1989}, Kleinian circle packings \cite{Mantica1995,Parker1995}, and random packings \cite{Baram2005}. This work is about space-filling packings as the one shown in Fig.~\ref{fig:RezaPacking}, which are generated using inversive geometry as in Refs.~\cite{Herrmann1990,Borkovec1994,Oron2000,Baram2004a,Stager2016a} and which therefore are exactly self-similar. Reference \cite{Baram2004a} previously showed that the packing in Fig.~\ref{fig:RezaPacking} is not a homogeneous fractal, since two planar cuts have different fractal dimensions; but no further investigation was carried out.

Here, we show that for all the self-similar space-filling packings constructed by inversive geometry of Refs.~\cite{Herrmann1990,Borkovec1994,Oron2000,Baram2004a,Stager2016a}, cuts along random hyperplanes generally have a fractal dimension of the one of the uncut packing minus one, what we prove analytically. Nevertheless, these packings are still heterogeneous fractals since cuts along special hyperplanes of a single packing show specific fractal dimensions. We present a strategy to search for such special cuts, which we illustrate on the packing in Fig.~\ref{fig:RezaPacking} as well as on a four-dimensional packing of Ref.~\cite{Stager2016a} out of which one can cut, for instance, the packing in Fig.~\ref{fig:RezaPacking}.

This study is organized in the following way. In Sec.~\ref{sec:properties_of_packings}, we introduce some general properties of the considered packings. In Sec.~\ref{sec:cutting}, we deal with cuts of packings where we consider random cuts in Sec.~\ref{sec:random_cuts}, and special cuts, as mentioned before, in Sec.~\ref{sec:special_cuts}. We draw conclusions in Sec.~\ref{sec:conclusion}.

\section{Properties of Packings}\label{sec:properties_of_packings}

\begin{figure}[t]
\begin{center}
	\includegraphics[width=\columnwidth]{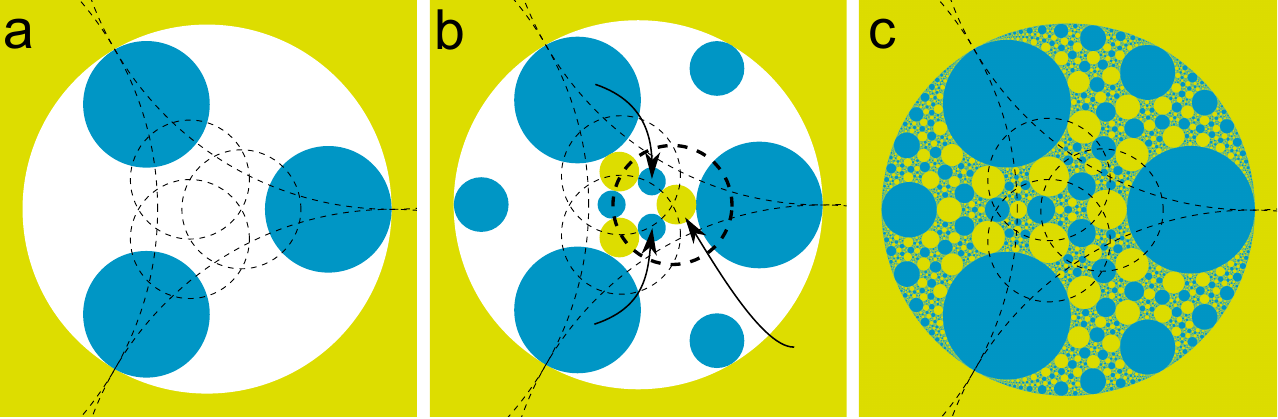}
\end{center}
\caption{
\label{fig:construction_principle} Principle of constructing a space-filling packing using inversive geometry: (a) Initially placed seeds (filled) are iteratively inverted at a group of inversion circles (dashed). (b) Packing after first iteration of inversions. Arrows point from seeds to their inverses with respect to the highlighted inversion circle. (c) Self-similar space-filling packing resulting after infinite iterations.
}
\end{figure}

We deal with space-filling packings that are constructed using inversive geometry, namely circle inversion, which leads to exactly self-similar packings. Since circle inversion is a conformal mapping, the inverse of a circle with respect to an inversion circle is again a circle. If the original circle lies outside the inversion circle, its inverse lies on the inside and is smaller. This principle is used to generate a space-filling packing from an initial set of disks called seeds, which are iteratively inverted at multiple inversion circles, as shown in Fig.~\ref{fig:construction_principle}.
With each iteration of inversions, more space is filled with smaller disks till eventually all space is covered, evidently leading to self-similar fractal packings. The final packing is invariant with respect to the inversion circles used for generating the packing, since they map the packing onto itself.

A more detailed explanation of the construction method can be found in Ref.~\cite{Stager2016a}, where we show how to generate, using inversive geometry, a variety of packings in two, three, and four dimensions, including all the topologies of Refs.~\cite{Herrmann1990,Borkovec1994,Oron2000,Baram2004a}. Circle inversion can be straightforwardly extended to sphere inversion in any higher dimension, such that everything we explain here in two or three dimensions holds analogously for higher dimensions.

\begin{figure}[]
\begin{center}
	\includegraphics[width=\columnwidth]{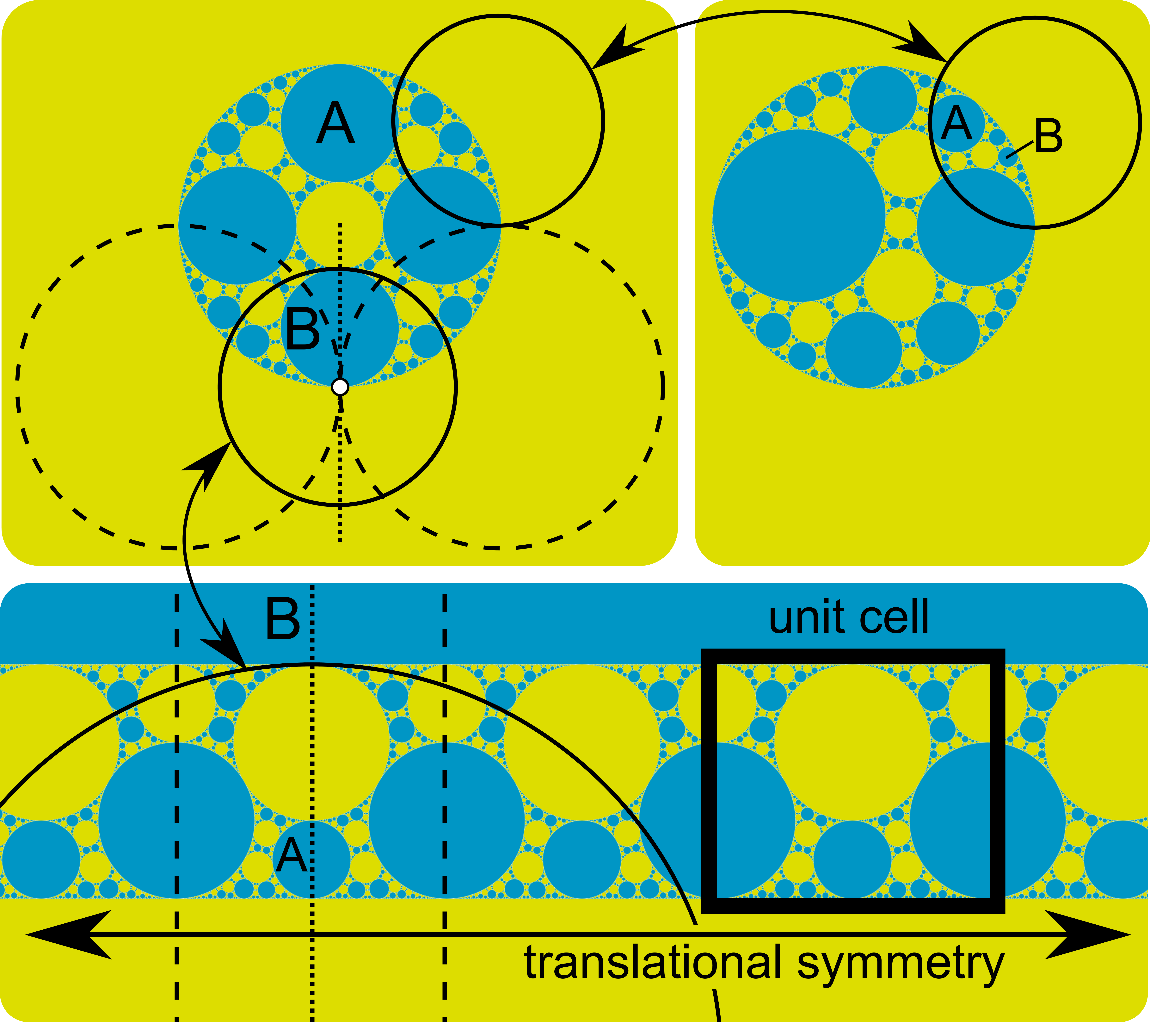}
\end{center}
\caption{
\label{fig:circle_to_strip} Inversion of a whole packing at different inversion circles (solid). A highly symmetric packing (top left) can be inverted onto an asymmetric one (top right). When the packing (top left) is inverted at an inversion circle with center at a contact point (white point) of two disks which also is a contact point of two inversion circles (dashed) with respect to which the packing is invariant, it is mapped onto a periodic structure enclosed by two parallel lines, i.e., a strip configuration (bottom). In this form the packing has a unit cell with translational symmetry.
}
\end{figure}

Apart from generating packings, circle inversions can also be used to invert a whole packing. That can change the sizes and spatial arrangement of its disks or spheres, but the topology and fractal dimension are invariant with respect to inversion. Figure \ref{fig:circle_to_strip} shows how one can invert a packing in different ways. For example, a highly symmetric packing can be mapped onto an asymmetric one. Furthermore, by inverting a 2D packing with respect to a circle whose center lies on a contact point between two disks, these two disks are mapped onto disks with an infinite radius, i.e., parallel lines that enclose the inverse of the packing. This comes from the fact that an inversion circle maps its own center onto infinity. This kind of configuration is called the strip configuration. The packing in between has a finite unit cell with periodic continuation, i.e., translational symmetry, as shown at the bottom of Fig.~\ref{fig:circle_to_strip}. The unit cell is bounded by two mirror lines which are the inverse of two tangent inversion circles with respect to which the packing is invariant. By construction, at every contact point of touching spheres, one can find two such inversion circles tangent to each other. Each of the packings considered here from Refs.~\cite{Herrmann1990,Borkovec1994,Oron2000,Baram2004a,Stager2016} can analogously be mapped onto a strip configuration, which for any considered dimensions is enclosed by two hyperplanes.

\section{Cutting}\label{sec:cutting}

\begin{figure}[]
\begin{center}
	\includegraphics[width=\columnwidth]{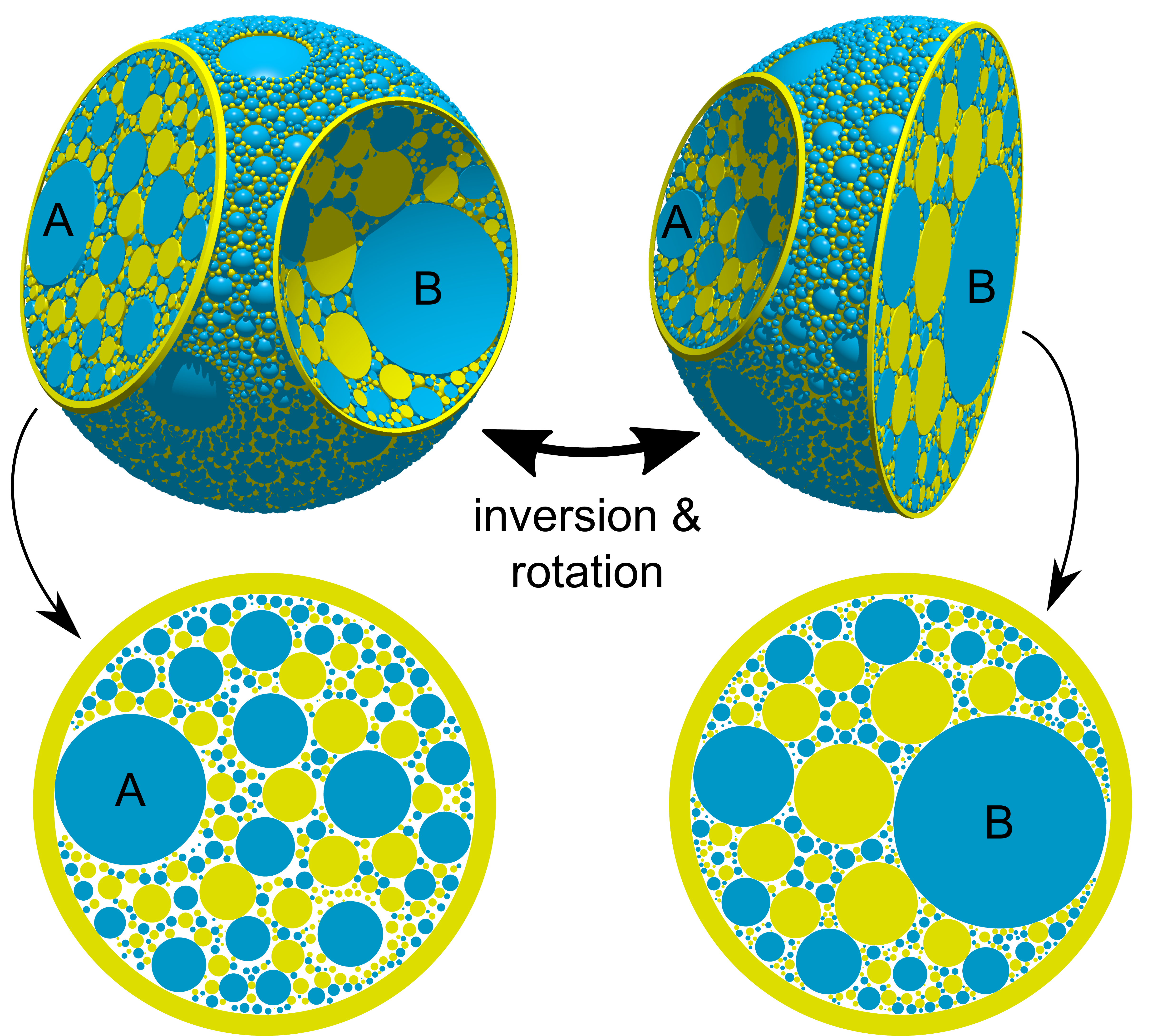}
\end{center}
\caption{
\label{fig:spherical_and_plane_cuts} A packing can be cut by any plane and more generally by any sphere. By inverting the whole packing one can map any spherical cut into a planar one and vice versa.
}
\end{figure}

To obtain different cuts of a sphere packing, one can cut along different planes. More generally, one can cut along any sphere, since by inverting the whole packing, one can transform every spherical cut into a planar one and vice versa, as shown in Fig.~\ref{fig:spherical_and_plane_cuts}.

\subsection{Random Cuts}\label{sec:random_cuts}

Analogously to Fig.~\ref{fig:circle_to_strip}, one can map a sphere packing onto a periodic strip configuration enclosed by two planes. We use this fact to derive the fractal dimension of random cuts, as explained in the following.

\begin{figure}[]
\begin{center}
	\includegraphics[width=\columnwidth]{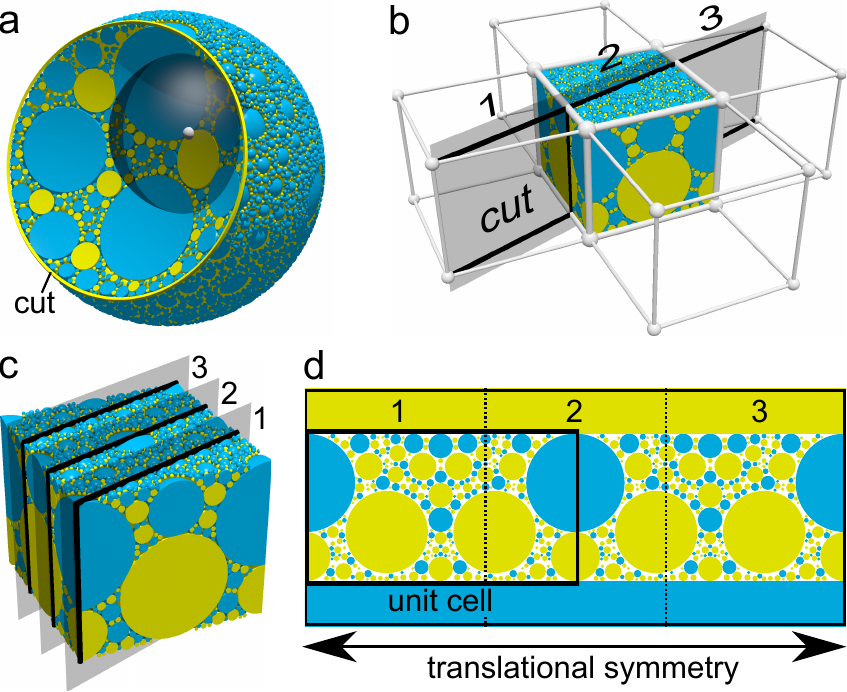}
\end{center}
\caption{
\label{fig:spherical_cut_to_plane_in_unit_cell} (a) Any spherical or planar cut of a packing can be inverted at an inversion sphere (transparent) with center (white) at the contact point between two touching cut spheres to map the cut onto a planar cut that is cutting the packing in its strip configuration as shown in (b). Different unit cells (1,2,3) of the strip configuration in (b) are cut at different positions. One can project all unit cells together with the cutting plane onto a single unit cell as shown in (c). In this particular case, this leads to only three different unit-cell cuts. The cut can be formed out of a sequence of the three unit-cell cuts resulting in a periodic strip configuration as shown in (d). Therefore, also the cut itself has a unit cell with translational symmetry. If the resulting cut is periodic or not, depends on the orientation of the cutting plane.
}
\end{figure}

\begin{figure}[]
\begin{center}
	\includegraphics[width=\columnwidth]{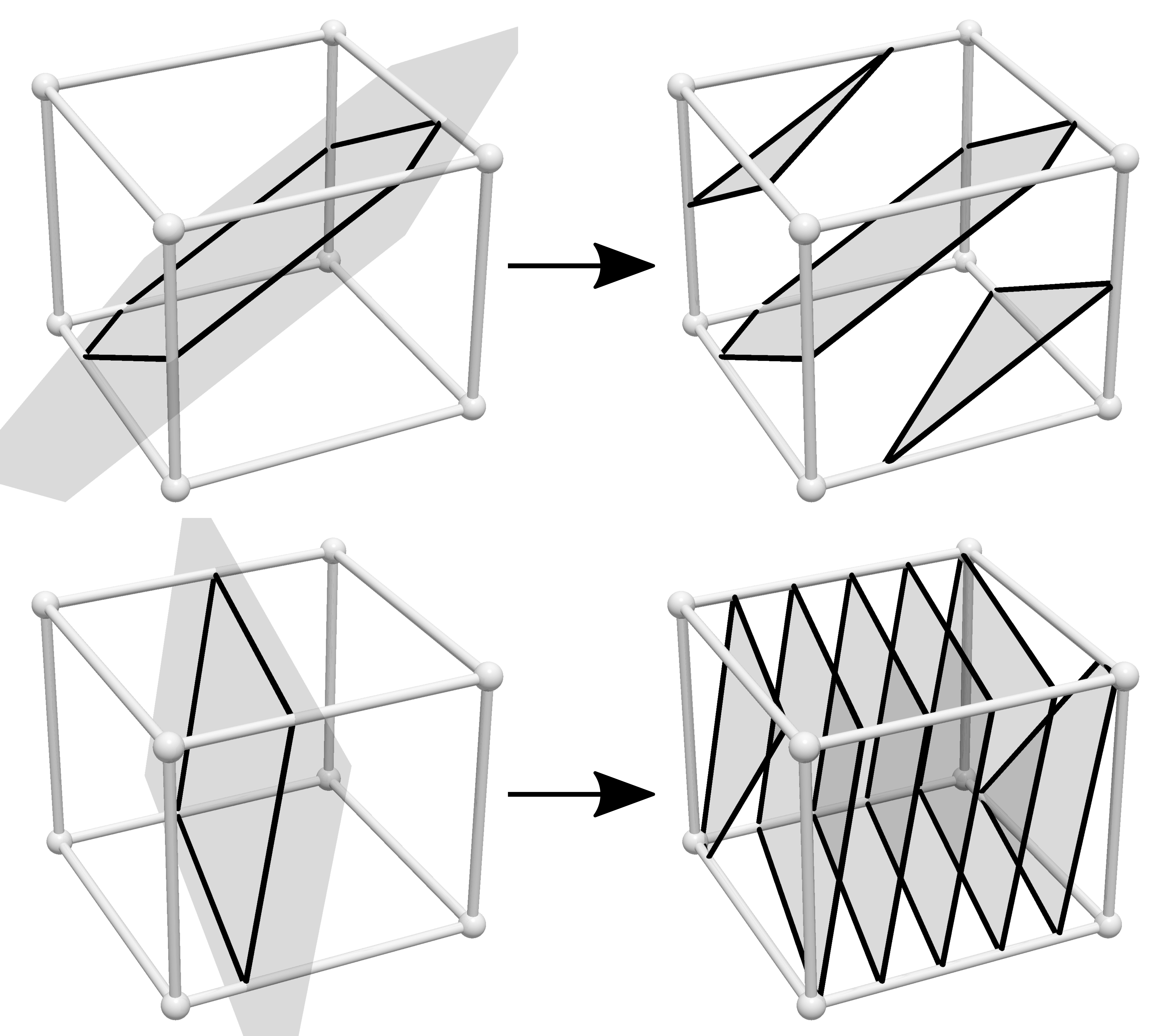}
\end{center}
\caption{
\label{fig:different_orientation_different_cut_sums} Different special orientations of cutting planes (left) for which the cut can be formed out of a finite number of different unit-cell cuts (right). In the general case of a random orientation of the cutting plane, infinitely many different unit-cell cuts need to be combined to form the cut as shown in Fig.~\ref{fig:random_orientation_homogeneous_cut_density}.
}
\end{figure}

\begin{figure}[]
\begin{center}
	\includegraphics[width=\columnwidth]{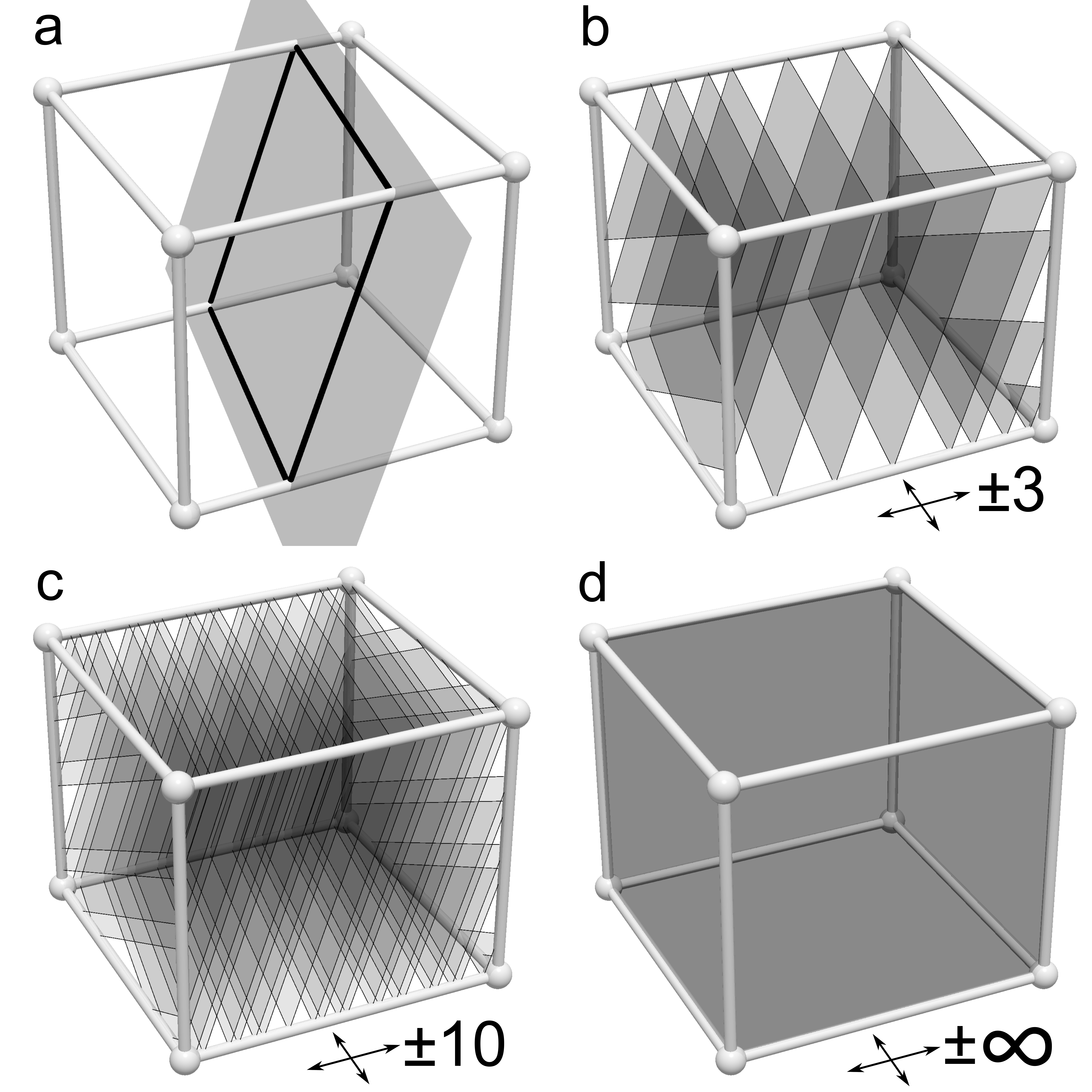}
\end{center}
\caption{
\label{fig:random_orientation_homogeneous_cut_density} A randomly oriented cutting plane can in general only be formed out of a combination of infinitely many different unit-cell cuts. (a) Randomly oriented cutting plane cutting a single unit cell. (b,c) Projection of neighboring unit cells together with the cutting plane onto a single unit cell, for different ranges of projection. (d) Projection of all unit cells leading to infinitely many different unit-cell cuts.
}
\end{figure}

Any cut, planar or spherical, can be mapped onto a planar cut in the strip configuration as shown in Fig.~\ref{fig:spherical_cut_to_plane_in_unit_cell}(a,b).
The strip configuration has a unit cell with translational symmetry in two dimensions parallel to the two planes that enclose the packing. In this periodic structure, we can look at each unit cell individually. We will generally find that some cells are cut by the cutting plane and some are not. Of the ones that are cut, each individual cell might be cut at a different position by the cutting plane. Let us project all unit cells together with the cutting plane onto a single unit cell as shown in Fig.~\ref{fig:spherical_cut_to_plane_in_unit_cell}(c). For the specifically chosen cut in Fig.~\ref{fig:spherical_cut_to_plane_in_unit_cell}, this projection results in three different unit-cell cuts. Therefore, the cut has a periodic structure since it can be formed out of a sequence of these three unit-cell cuts, infinitely repeating itself. Depending on the orientation of the initial cut, the projection results in a different number of unit-cell cuts as shown in Fig.~\ref{fig:different_orientation_different_cut_sums}.
If the cut is chosen randomly, it results generally in an infinite number of different unit-cell cuts, as indicated in Fig.~\ref{fig:random_orientation_homogeneous_cut_density}.
In principle, this is the same as using the well-known "cut-and-project" method to obtain a quasiperiodic structure as an "irrational slice" of a periodic lattice \cite{Janot1992,Steurer1999,Baake2002,Man2005,Ledermann2006,Moretti2007,Rodriguez2008}. Important here is that the density of unit-cell cuts is homogeneous across the whole unit cell. Out of a single sphere of the unit cell, infinitely many disks are cut. In detail, one finds the density of disks of radius $r$ that are cut out of spheres of radius $R$ to be
\begin{equation}
\label{eq:n2D_partial}
\tilde{n}_\text{cut}(r,R) =
  \begin{cases}
    \frac{2r}{\sqrt{R^2-r^2}} n(R)    & \quad \text{for }0 < r \leq R \text{,}\\
    0  & \quad \text{otherwise,}\\
  \end{cases}
\end{equation}
where $n(R)$ is the density of spheres or radius $R$. When we consider spheres of all radii, a disk of radius $r$ can be a cut out of any sphere of radius $R \geq r$. Therefore, we can obtain the density $n_\text{cut}(r)$ of disks in the cut by considering all spheres of radius $R \geq r$. We integrate the density $\tilde{n}_\text{cut}(r,R)$ of disks that are cut out of spheres of radius $R$, which we defined in Eq.~(\ref{eq:n2D_partial}), over all $R \geq r$ to find
\begin{equation}
\label{eq:n2D_full}
n_\text{cut}(r) = \int_r^\infty \tilde{n}_\text{cut}(r,R)\,\mathrm{d}R = \int_r^\infty \frac{2r}{\sqrt{R^2-r^2}} n(R)\,\mathrm{d}R.
\end{equation}
The density $n(R)$ of spheres follows asymptotically a simple power law
\begin{equation}
\label{eq:asymptotic_behavior_n}
n(R) \sim R^{-d_f-1},
\end{equation}
where $d_f$ is the fractal dimension of the packing \cite{Boyd1973a,Boyd1982,Mandelbrot1982,Manna1991,Borkovec1994}. From this we assume $n(R) = k \cdot R^{-d_f-1}$, where $k>0$ is a constant. Thus, we find from Eq.~(\ref{eq:n2D_full}) that  
\begin{equation}
\label{eq:n2D_final}
n_\text{cut}(r) = \int_r^\infty \frac{2kr}{\sqrt{R^2-r^2}} R^{-d_f-1}\,\mathrm{d}R=2\pi k \frac{\Gamma(\frac{d_f+1}{2})}{\Gamma(\frac{d_f}{2})}r^{-d_f},
\end{equation}
where $\Gamma$ denotes the gamma function with $\Gamma(t)=\int_0^\infty x^{t-1} e^{-x} \, \mathrm{d}x$. In Eq.~(\ref{eq:n2D_final}), we see that $n_\text{cut}(r) \sim r^{-d_f}$ and we know from Eq.~(\ref{eq:asymptotic_behavior_n}) that $n_\text{cut}(r) \sim r^{-d_{f\text{,cut}}-1}$, where $d_{f\text{,cut}}$ is the fractal dimension of the cut. Therefore, we find $d_{f\text{,cut}}=d_f-1$, i.e., the fractal dimension of random cuts is always the one of the uncut packing minus one.

\subsection{Special Cuts}\label{sec:special_cuts}

To generate a packing we use seeds and inversion circles which together are called the generating setup \cite{Stager2016a}. Different topologies originate from different generating setups. Nevertheless, some setups lead to the same packing and some generate different packings but the same topology \cite{Stager2016a}. In the latter case, the packings can be mapped onto each other through a certain sequence of inversions. However, since we are interested in finding special cuts with distinct fractal dimensions in a single packing, we will look for cuts with different generating setups.

Let us first describe how one can find a generating setup of a cut. Like the one shown in Fig.~\ref{fig:construction_principle}a, every generating setup consists of seeds and inversion circles. No seed lies completely inside an inversion circle, such that all its inverses are smaller. To lead to a space-filling packing, the seeds and inversion circles together need to cover all space, as proven in Ref.~\cite{Stager2016a}. In a cut of a sphere packing, every disk is a potential seed for a generating setup. Additionally, we need potential inversion circles, which we find as shown in the following.

\begin{figure}[]
\begin{center}
	\includegraphics[width=\columnwidth]{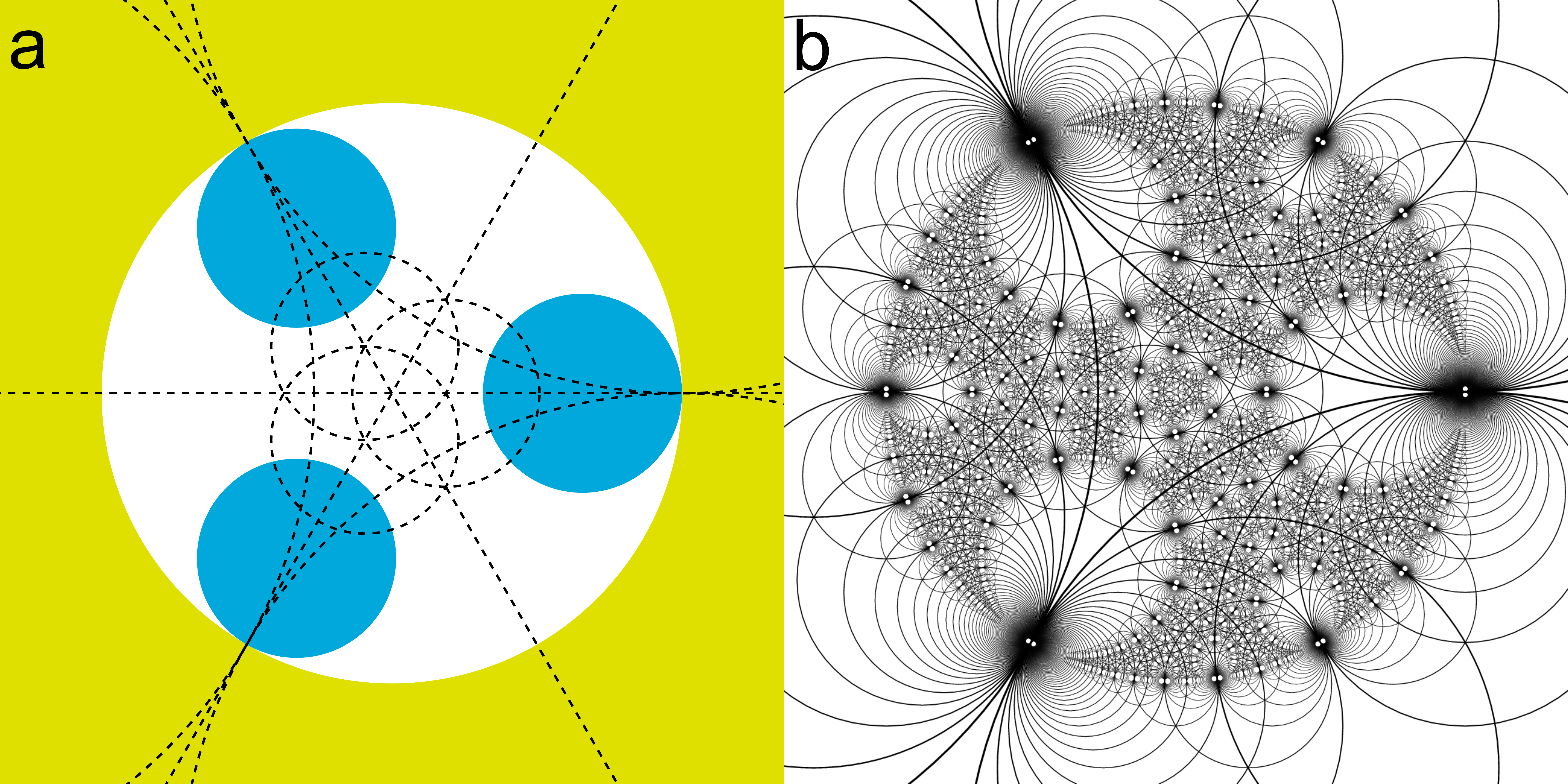}
\end{center}
\caption{
\label{fig:all_invariant_circles} (a) Seeds (filled) with inversion circles and mirror lines (dashed). (b) All inversion circles with respect to whom the packing is equal to its inverse. Obtained from iterative inversions of the inversion circles and mirror lines in (a) with respect to each other.
}
\end{figure}

\begin{figure}[]
\begin{center}
	\includegraphics[width=\columnwidth]{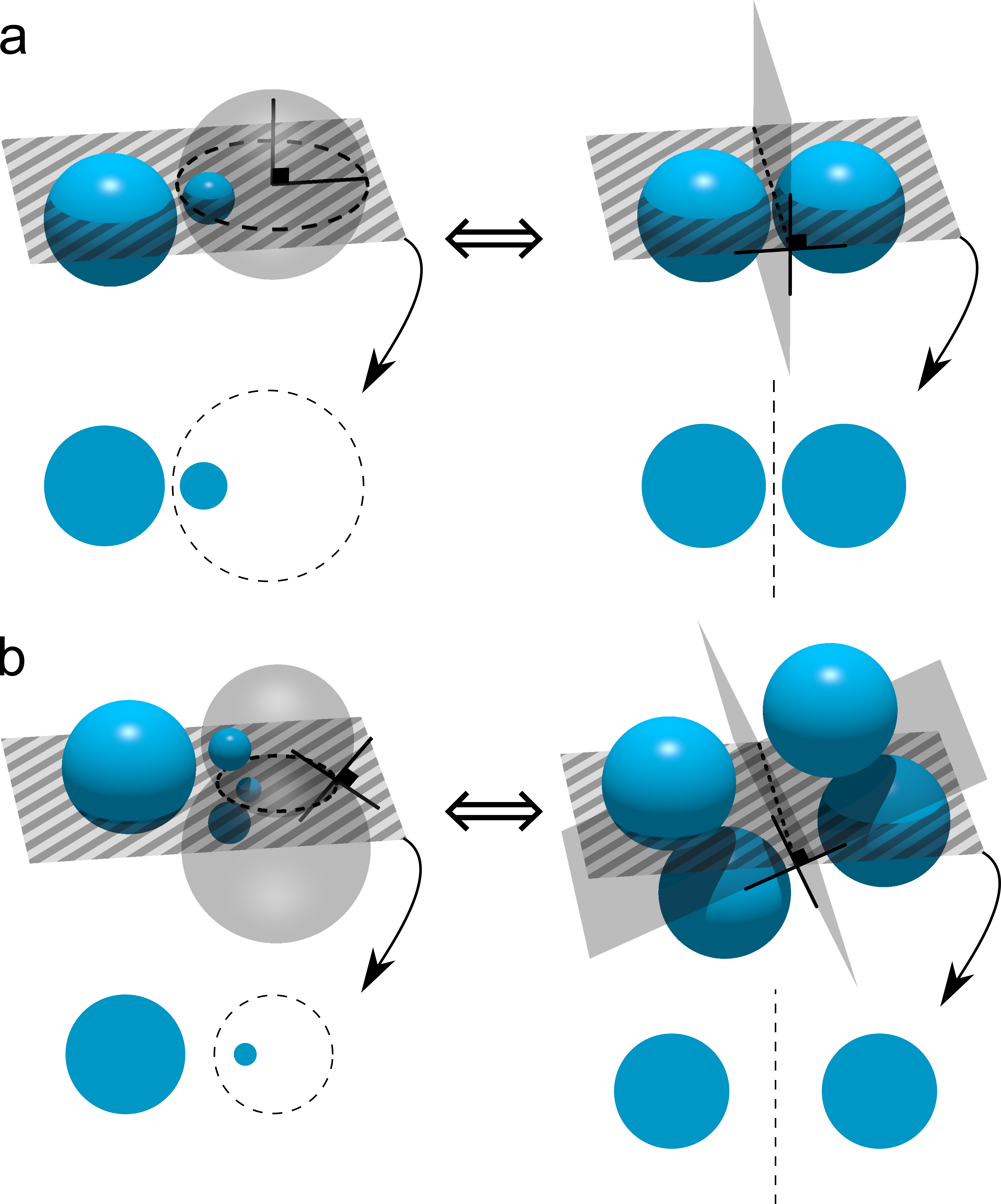}
\end{center}
\caption{
\label{fig:two_ways_how_inversion_circles_appear} Inversion circles (dashed circles) in the cut (dashed plane) are either: (a) (left) Cuts perpendicular to inversion spheres (transparent) what is topologically the same as (right) a cut perpendicular to a mirror plane (transparent); or (b) (left) Cuts containing the intersection of two inversion spheres (transparent) that intersect each other perpendicularly what is topologically the same as (right) a cut through the intersection line of two mirror planes (transparent) perpendicular to each other. Disks are just the cuts of spheres.
}
\end{figure}

From the generating setup of the sphere packing, we can first derive all inversion spheres with respect to which the packing is invariant. As shown on a two-dimensional example in Fig.~\ref{fig:all_invariant_circles}, all inversion spheres together with all mirror planes of the generating setup need to be iteratively inverted at each other to find all inversion spheres with respect to which the packing is invariant.
Given these inversion spheres of the packing, we can derive the inversion circles of the cut. Unlike every cut of a sphere results in a disk, not every cut of an inversion sphere is also an inversion circle in the cut. Considering a single inversion sphere, only when cutting perpendicularly to it, the resulting circle is an inversion circle in the cut as shown in Fig.~\ref{fig:two_ways_how_inversion_circles_appear}a.
This is topologically the same as that only when one cuts perpendicularly to a mirror plane, the intersection line is a mirror line in the cut. Considering multiple inversion spheres, there is one additional way how an inversion circle can appear in a cut. Any circle, which is the intersection of two inversion spheres that intersect each other perpendicularly, is an inversion circle in the cut as shown in Fig.~\ref{fig:two_ways_how_inversion_circles_appear}b. That is topologically the same as that the intersection line of two mirror planes perpendicular to each other turns out to be a mirror line in the cut. Apart from the here derived inversion circles which are cuts of inversion spheres, there could in principle also appear inversion circles in a cut which do not lie on the surface of any inversion sphere, but for simplicity, we neglected this more complex scenario.

After having found all disks and inversion circles of a cut, one needs to check if they together cover all space. If they do, one can, to end up with a generating setup, neglect every disk and inversion circle whose center is inside another inversion circle. These disks and inversion circles are redundant since they can be generated from a larger disk and inversion circle, respectively, by inverting at the inversion circle in which their center lies.

For a given smallest radius of spheres and inversion spheres of a packing, one can for a random cut generally not find a generating setup, because in general one finds no inversion circles in a random cut, which one would need to find to be able to cover the empty space between the disks. To find cuts that we can generate, we use the following strategy. We first find all inversion spheres larger than the radius $r_\text{find}$. We want to find cuts in which the cuts of these inversion spheres appear as possible inversion circles (compare Fig.~\ref{fig:two_ways_how_inversion_circles_appear}a). We divide the inversion spheres into the ones that intersect the unit sphere (outer inversion spheres) and that do not (inner inversion spheres). We then find all spherical cuts that are perpendicular to three outer inversion spheres and one inner, and all planar cuts that are perpendicular to three outers. We chose this strategy for its computational efficiency since every special cut needs to contain at least three outer and one inner inversion circle, and there are many more inner than outer inversion spheres. 

Since some cuts lead to the same topology, we rule some multiple appearances out the following way. We only consider cuts that have no inverses larger than themselves and whose centers lie in a chosen area bounded by mirror planes of the packing, where the center of a planar cut lies at infinity in the direction of its normal vector. Some topologies might still appear multiple times, which cannot be mapped onto each other with a single inversion, such that one has to sort them out separately. This comes from the fact that for the particular packing considered here, one can even map the spheres directly touching each other onto each other not by a single inversion but by a sequence of multiple inversions. For all pairs of topologies with overlapping confidence intervals of the fractal dimensions, determined numerically as described later, we therefore made a topological comparison, as explained in detail in Appendix~\ref{app:topological_comparison}, to judge if they are different topologies or not.

\begin{figure}[t]
\begin{center}
	\includegraphics[width=\columnwidth]{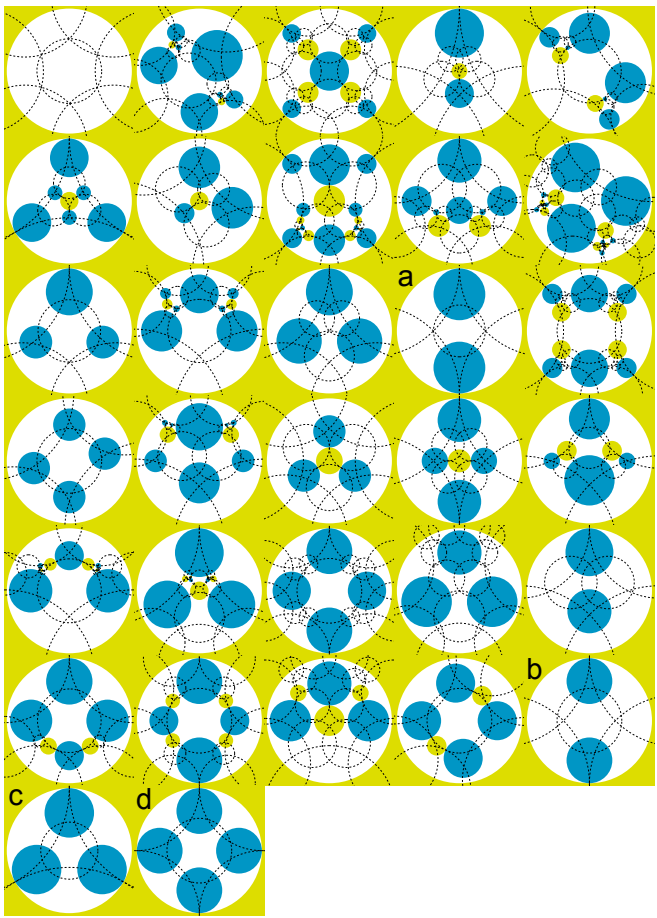}
\end{center}
\caption{
\label{fig:2D_32Cuts} Rescaled generating setups of special cuts out of the packing shown in Fig.~\ref{fig:RezaPacking}. They are ordered according to their fractal dimension decreasing from left to right and top to bottom. Topologies marked with a letter were already previously discovered by Ref.~\cite{Herrmann1990}, according to which they have the following parameters in the form (\emph{family},\emph{m},\emph{n}): (a) (2,1,1), (b) (2,0,1), (c) (1,0,0), and (d) (1,1,1).
}
\end{figure}

\begin{figure}[]
\begin{center}
	\includegraphics[width=\columnwidth]{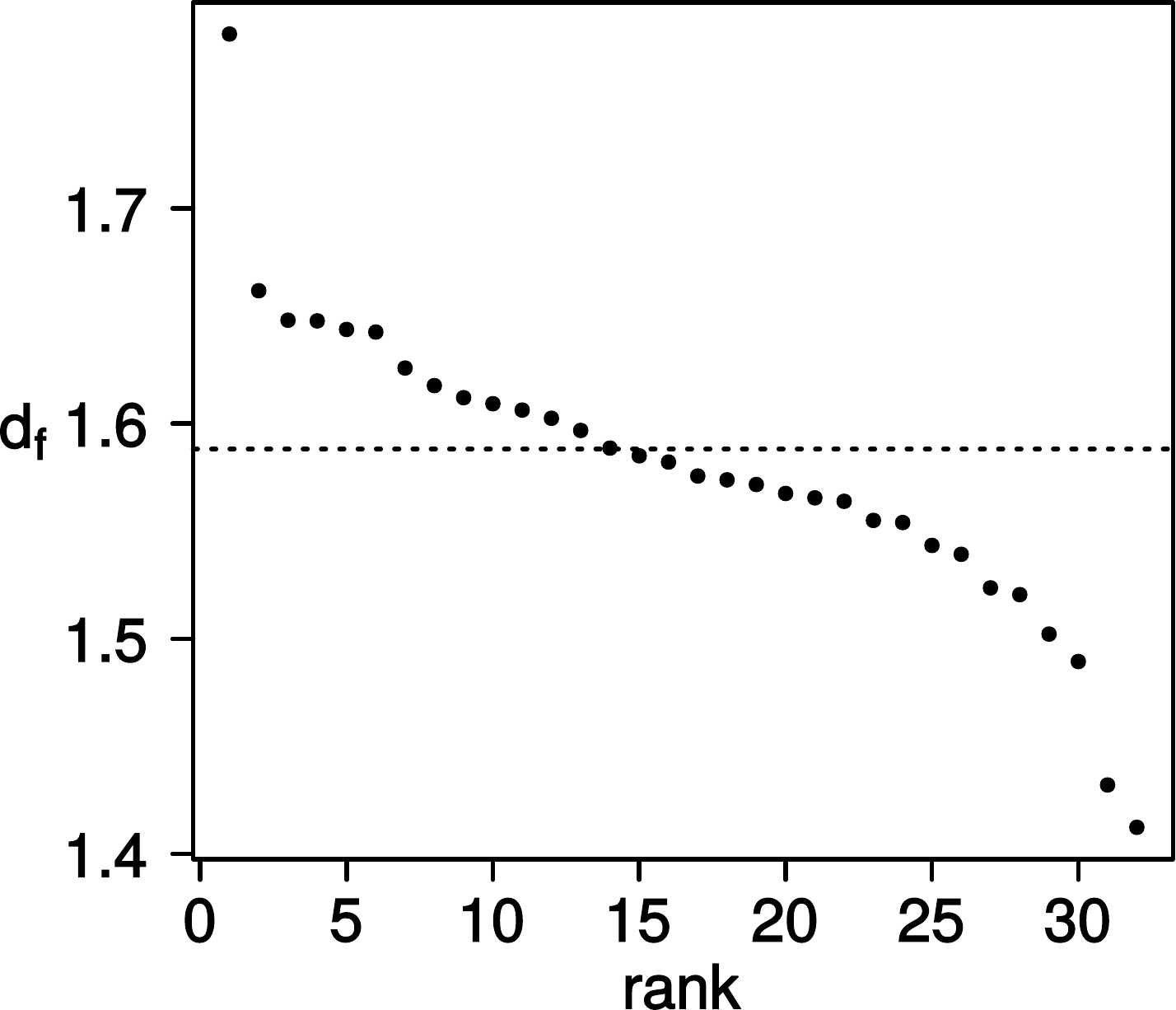}
\end{center}
\caption{
\label{fig:2D_32Cuts_dfs} Ranked fractal dimensions of special cuts shown in Fig.~\ref{fig:2D_32Cuts}. The dashed line indicates the fractal dimension of random cuts in general, which is one less than the fractal dimension of the uncut packing. Confidence intervals are smaller than the symbol size.
}
\end{figure}

\begin{figure}[h]
\begin{center}
	\includegraphics[width=\columnwidth]{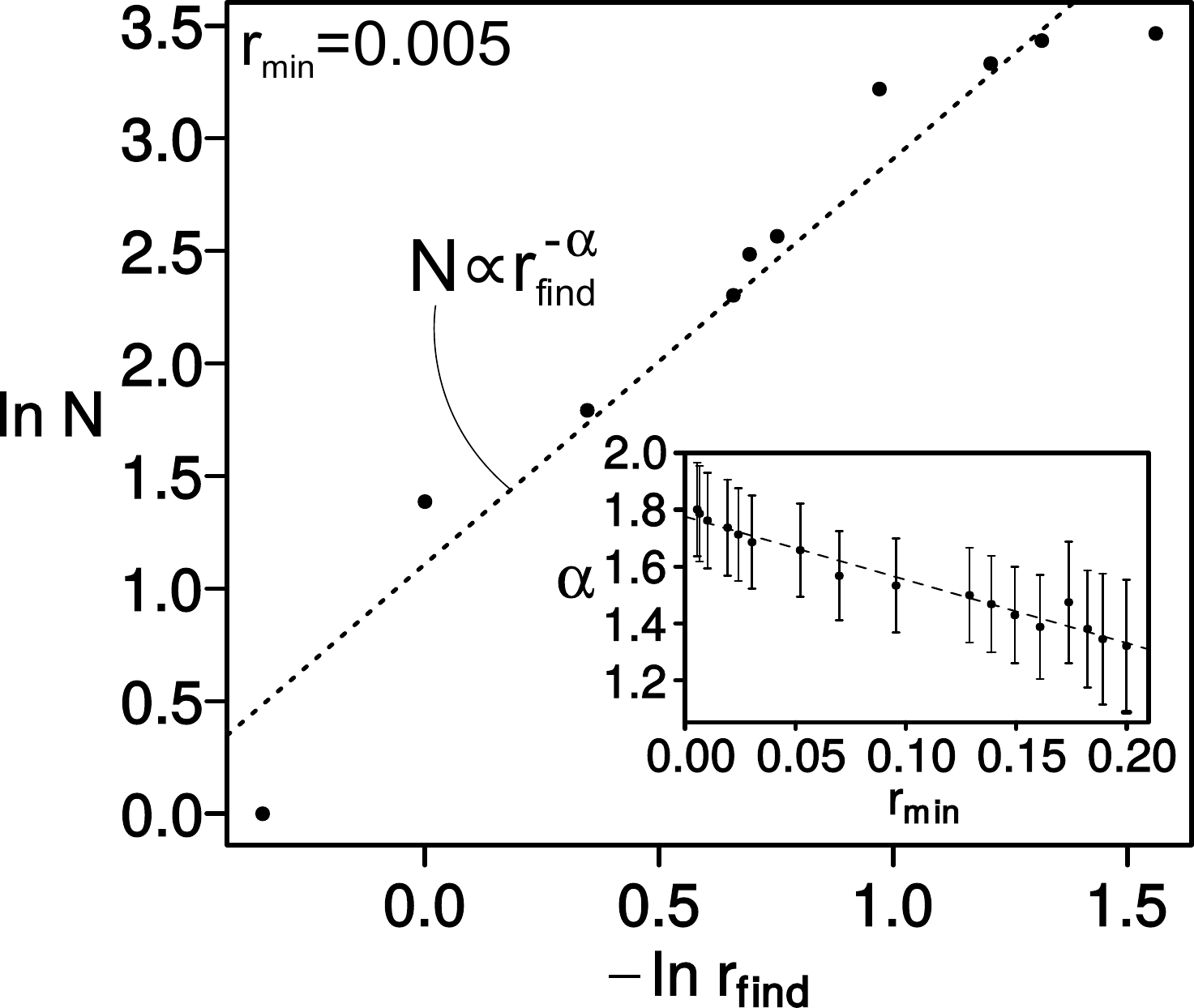}
\end{center}
\caption{
\label{fig:number_vs_rmin} The number of different found topologies $N$ versus the smallest considered radius $r_\text{find}$ of inversion spheres to define the cut for smallest considered cut spheres and inversion spheres of radius $r_\text{min}=0.005$. We find that $N$ can be approximately described by a power law $N \propto r_\text{find}^{-\alpha}$. Inset: estimates of the exponent $\alpha$ versus different $r_\text{min}$. We assume the estimated $\alpha$ is linearly dependent on $r_\text{min}$ such that we predict for the limiting case of $r_\text{min} \to 0$ that $\alpha =1.78 \pm 0.16$. This suggests that in the limit of $r_\text{min} \to 0$, one can find infinitely many different topologies for $r_\text{find} \to 0$.
}
\end{figure}

Using the described strategy, we searched for special cuts in the packing shown in Fig.~\ref{fig:RezaPacking}. We generated the packing down to a smallest radius of spheres and inversion spheres with respect to which the packing is invariant (compare Fig.~\ref{fig:all_invariant_circles}) of $r_\text{min}=0.005$. For the smallest radius of inversion spheres considered to define the cutting sphere or plane, we chose $r_\text{find} = 0.2$. We found 32 special cuts resembling different topologies. Their rescaled generating setups can be found in Fig.~\ref{fig:2D_32Cuts}, and their fractal dimensions in Fig.~\ref{fig:2D_32Cuts_dfs}.
The fractal dimensions were determined as in Ref.~\cite{Stager2016a}, considering the packings with all disks of radius larger than $e^{-12}$. We plot the number of different found topologies versus the search cutoff radius $r_\text{find}$ in Fig.~\ref{fig:number_vs_rmin}.
The number of found topologies $N$ seems to follow a power law $N \propto r_\text{find}^{-\alpha}$. For cuts for which we did not find a generating setup, we cannot be sure if they are special cuts or not, since we only cut spheres and inversion spheres larger than $r_\text{min}=0.005$. Considering a smaller $r_\text{min}$, one might find a generating setup. We therefore estimate $\alpha$ for different $r_\text{min}$ to make a prediction for the limiting case $r_\text{min} \to 0$. We assume the estimated $\alpha$ is linearly dependent on $r_\text{min}$ as shown in the inset of Fig.~\ref{fig:number_vs_rmin}. We conclude for $r_\text{min} \to 0$ that $\alpha = 1.78 \pm 0.16$. This suggests that one will find an infinite amount of special cuts corresponding to different topologies in the limit of $r_\text{find} \to 0$ and $r_\text{min} \to 0$. We assume this to be true for any three and higher-dimensional packing.

\begin{figure}[b]
\begin{center}
	\includegraphics[width=\columnwidth]{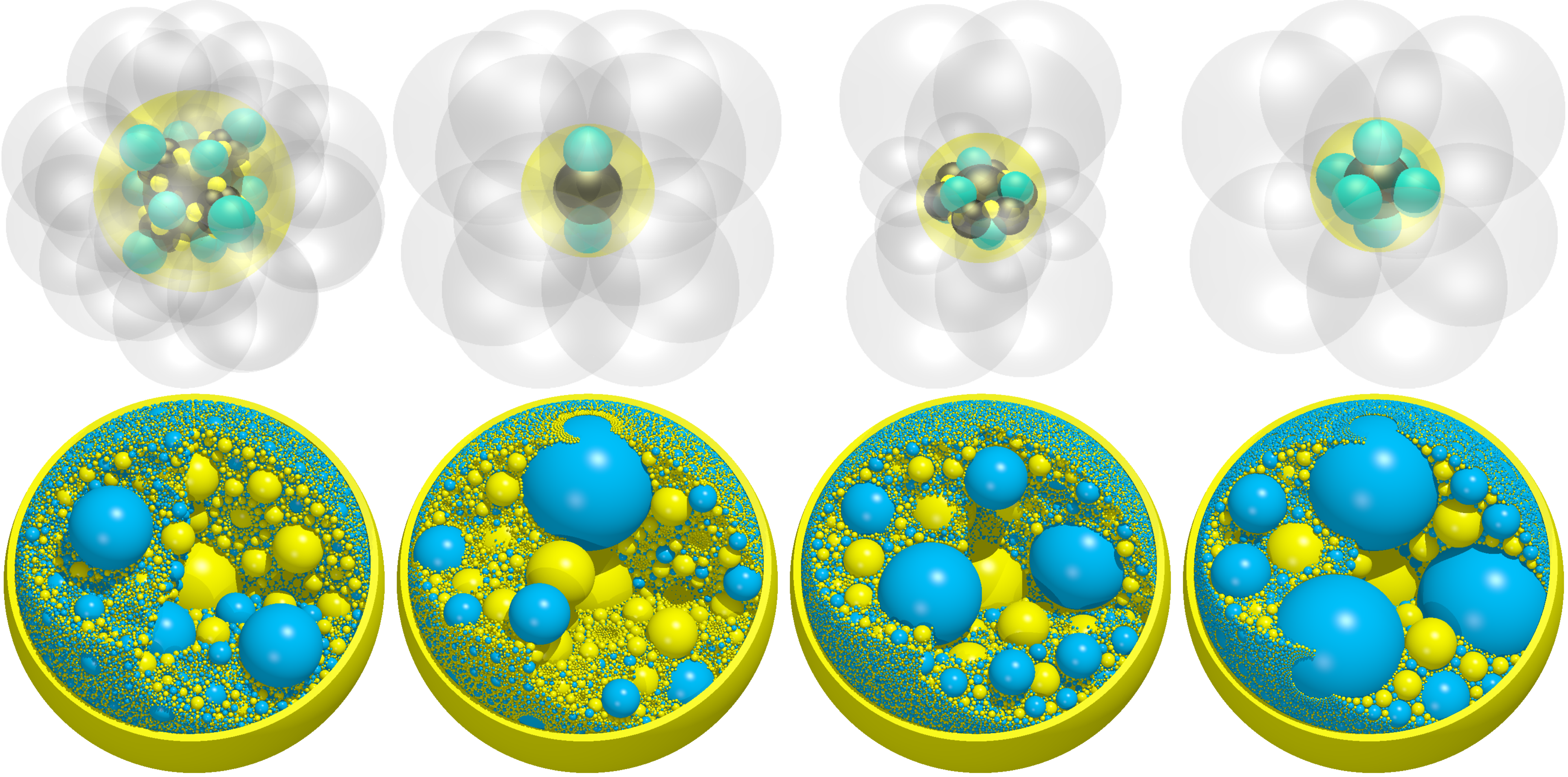}
\end{center}
\caption{
\label{fig:3D_Cuts} Generating setups (top) of special cuts (bottom) out of a four-dimensional packing of Ref.~\cite{Stager2016a}. (top) Inversion spheres in light grey and black and seeds in color. (bottom) Some spheres are removed to allow looking inside the three-dimensional cuts. The fractal dimension of the four-dimensional packing is $3.70695 \pm 0.0003$. The ones of the three-dimensional cuts are from left to right $2.780581 \pm 0.000003$, $2.735424 \pm 0.000005$, $2.70812 \pm 0.00002$, and $2.588191 \pm 0.000005$, where the first and third cut are newly discovered topologies whereas the second is previously known from Ref.~\cite{Stager2016a} and the last one is the packing in Fig.~\ref{fig:RezaPacking}.
}
\end{figure}

To demonstrate that our cut strategy can analogously be applied to higher-dimensional packings, we also cut a four-dimensional packing of Ref.~\cite{Stager2016a}. According to the definition and nomenclature in Ref.~\cite{Stager2016a}, this packing can be constructed from a generating setup based on the 16-cell with outer inversion spheres placed in the direction of its faces and seeds in the direction of its vertices, which we choose to lie at $(\pm1,0,0,0)$ and its permutations, and belonging to \emph{family} $1$ with parameters $b=c=0$. Its fractal dimension is $3.70695 \pm 0.0003$ \cite{Stager2016a}, such that for random cuts one would find a fractal dimension of one less in the range of $2.70695 \pm 0.0003$. For $r_\text{find}=0.5$ and $r_\text{min}=0.05$, we found four special cuts which are shown in Fig.~\ref{fig:3D_Cuts} together with their generating setups, which we order according to their fractal dimensions which we found to be $2.780581 \pm 0.000003$, $2.735424 \pm 0.000005$, $2.70812 \pm 0.00002$, and $2.588191 \pm 0.000005$. All are planar cuts through the center of the packing with normal vectors along $(1,1,1,1)$,$(1,1,1,0)$,$(2,1,1,0)$, and $(1,0,0,0)$, respectively.
The second and the last cut are topologies discovered previously and the first and third cut are new discoveries. The last cut is exactly the packing of Fig.~\ref{fig:RezaPacking}. The second one is the topology that according to Ref.~\cite{Stager2016a} can, for instance, be constructed from a generating setup based on the tetrahedron with outer inversion spheres in the direction of its faces belonging to \emph{family} $2$ with parameters $b=0$ and $c=1$. Due to the high computational effort needed to find special cuts in a four-dimensional packing, we chose a relatively large radius $r_\text{find}=0.5$ for the smallest inversion spheres considered to define our cuts. Even though we only found four special cuts in this case, we expect for any four- and even higher-dimensional packing to find, analogously to the three-dimensional example before, an infinite number of special cuts, which correspond to different topologies, in the limit of $r_\text{find} \to 0$ and $r_\text{min} \to 0$.

\section{conclusion}\label{sec:conclusion}
We have shown that self-similar space-filling sphere packings created by inversive geometry as in Refs.~\cite{Herrmann1990,Borkovec1994,Oron2000,Baram2004a,Stager2016a} are inhomogeneous fractals but that random cuts generally have a fractal dimension of the one of the packing minus one. We presented a strategy to look for special cuts with distinct fractal dimensions which allows identifying many different topologies out of a single packing. Our numerical analysis suggests that in the limit of a vanishing cutoff of smallest considered radii, one can find infinitely many special cuts corresponding to different topologies. This allows using packings in higher than three dimensions to find new two and three-dimensional topologies, whose direct construction setup is far from being trivial.

Reference \cite{Baram2004a} previously found two planar cuts of the packing in Fig.~\ref{fig:RezaPacking} with different fractal dimensions, without further investigation. After the detailed analysis here, we know that one of these two cuts is the special cut shown in Fig.~\ref{fig:2D_32Cuts}(d), and the other one is a cut with the fractal dimension of the uncut packing minus one, which can not be constructed itself by inversive geometry.

Bipartite sphere packings like the one in Fig.~\ref{fig:RezaPacking} have drawn attention since they allow all spheres to rotate simultaneously in a specific way without any slip between neighboring spheres as shown in Refs.~\cite{Baram2004,Stager2016,Araujo2013}, such that some packings even allow the prediction and control of the slip-free rotation state \cite{Stager2016}. Regarding bipartite packings, it is still unknown if some can be used as a bearing to decouple the motion of two parallel planes as suggested in Ref.~\cite{Stager2016a}. By cutting four-dimensional bipartite topologies, one might be able to find three-dimensional sphere packings with previously unknown mechanical functionalities regarding their slip-free rotation state.

\begin{acknowledgments}
We acknowledge financial support from the ETH Risk Center, the Brazilian institute INCT-SC, Grant No.~FP7-319968-FlowCCS of the European Research Council (ERC) Advanced Grant.
\end{acknowledgments}

\bibliography{bibliography_paper}
\newpage
\appendix
\section{Topological comparison}\label{app:topological_comparison}

To judge if two different packings are the same topology or not, one can carry out a topological comparison. If two generating setups of the different packings are topologically equal, the packings are the same topology. Since a single packing can have different generating setups, one first needs to define a type of setup that is topologically unique for the resulting topology. We define a topologically unique setup as the minimal generating setup of a packing, i.e., the setup with the least amount of seeds and inversion spheres needed. 

\begin{figure}[]
\begin{center}
	\includegraphics[width=\columnwidth]{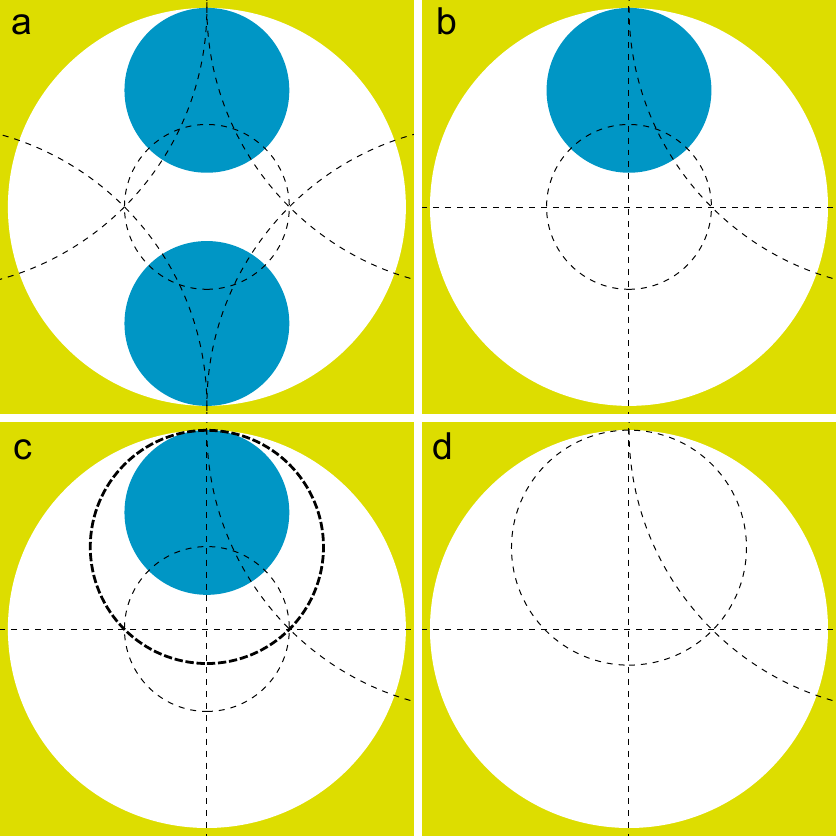}
\end{center}
\caption{
\label{fig:reduce_topology} Different steps from an original generating setup (a) to a minimal one (d). (b) reduced setup after considering mirror lines, i.e., infinitely large inversion circles. (c) Potential inversion circle (dashed, highlighted) that can be found from the fact that it maps the two seeds onto each other.
}
\end{figure}

Therefore, we first find from the generating setup of each packing a minimal setup. We explain this procedure at the two-dimensional example in Fig.~\ref{fig:reduce_topology}, which can be analogously applied to any higher dimension.

We start with the original setup (Fig.~\ref{fig:reduce_topology}a) and first find all mirror lines, which in a setup can be used as inversion circles of infinite radii. Seeds are only allowed to intersect inversion circles, including mirror lines, perpendicularly, otherwise, they have to lie outside of them. Therefore, we need to define which side of the mirror lines we consider as the outside. We choose an arbitrary point $P$ in space that we declare to lie outside of all mirror lines, where $P$ should not lie on a mirror line itself. We then neglect every inversion circle and seed that lie inside a mirror line. As shown in Fig.~\ref{fig:reduce_topology}b, this already leads to a reduced setup. From there, we check if any two inversion circles or any two seeds can be mapped onto each other by a new inversion circle. If we find such an inversion circle as shown in Fig.~\ref{fig:reduce_topology}c, we add it and iteratively invert every seed and inversion circle at inversion circles who they intersect with an angle larger than $\pi/2$. We do this to find the largest inverses of each seed and inversion circle which lies outside all mirror lines.

In the resulting setup, certain inversion circles and seeds might exist multiple times, such that we only keep a single instance of it. We check if this setup is valid, i.e., if it fulfills all constraints as discussed in detail in Ref.~\cite{Stager2016a}. If it is a valid setup, one needs to proof that it leads to the same packing as the original setup. One can do this by inverting every seed and inversion circle of the original setup iteratively at the inversion circles of the newly proposed setup, till one found the largest inverse of each original seed and inversion circle which lays outside of all mirror lines. If every of these largest inverses is equal to a seed or inversion circle of the newly proposed setup, respectively, we know that the original setup can be generated from the newly proposed one. Therefore, the newly proposed setup leads to the same packing as the original one. One needs to continue to try to reduce every newly accepted setup the same way, till one cannot minimize it any further, to be sure to have found the minimal setup, as the one shown in Fig.~\ref{fig:reduce_topology}d.

\begin{figure}[t]
\begin{center}
	\includegraphics[width=\columnwidth]{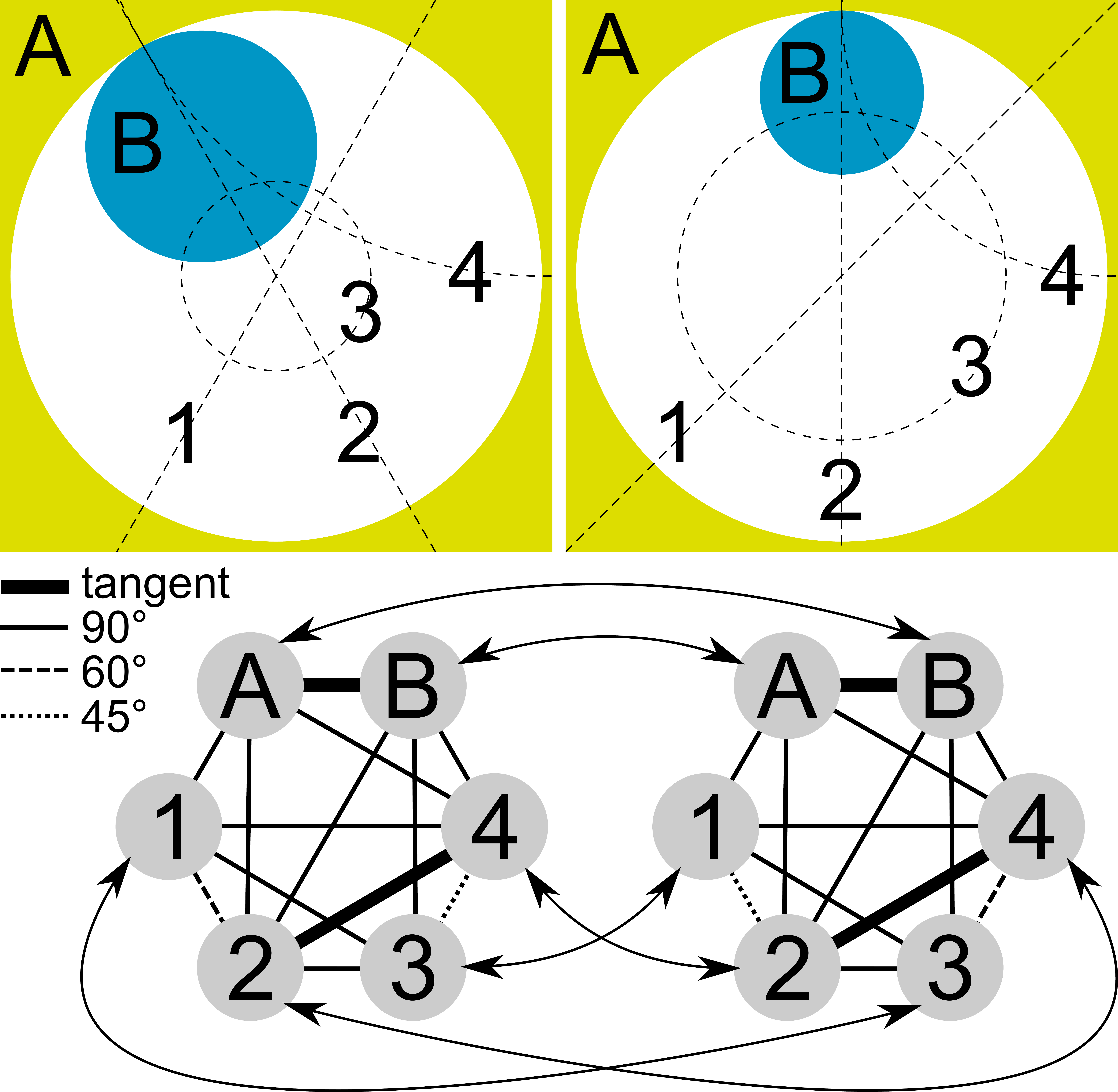}
\end{center}
\caption{
\label{fig:topological_comparison}
Two minimal generating setups (top) are topologically equivalent if one can find a bijection (arrows) between the two connection networks (bottom) of their seeds and inversion circles. Seeds and inversion circles are connected if they are tangent or if they intersect. The details of the connection, i.e., the intersection angle or the fact that they are tangent, can be seen as a weight or label of the connection.
}
\end{figure}
After having found two minimal setups for two different packings, one can topologically compare them. The whole topological information lies in the arrangement of the seeds and inversion circles, i.e., in the way the touch and overlap each other. The seeds and inversion circles form a network, where two elements are connected if they intersect or if they are tangent to each other. The kind of connection, i.e., the intersection angle or the fact that they are tangent, can be seen as a weight or label of the connection. If one can find a bijection between the two networks as shown in Fig.~\ref{fig:topological_comparison}, the two packings are the same topology, otherwise, they are different.

\end{document}